\documentclass[particles,article,accept,pdftex,moreauthors]{Definitions/mdpi} 

\firstpage{1} 
\pubvolume{1}
\issuenum{1}
\articlenumber{0}
\pubyear{2024}
\copyrightyear{2024}
\externaleditor{Academic Editor: Armen Sedrakian}
\datereceived{1 December 2023} 
\daterevised{23 December 2023} % Comment out if no revised date
\dateaccepted{8 January 2024} 
\datepublished{ } 
\hreflink{https://doi.org/} % If needed use \linebreak
\Title{Towards Uncovering Dark Matter Effects on Neutron Star Properties: A Machine Learning Approach} %MDPI: Title is different from our system. Please check and revise if necessary. %NOTE: Change made to title
\TitleCitation{Towards Uncovering Dark Matter Effects on Neutron Star Properties: A Machine Learning Approach} %NOTE: Citation updated to match title
\Author{Prashant %MDPI: Please carefully check the accuracy of names and affiliations. %TM: Confirmed.
 Thakur $^{1}$, Tuhin Malik $^{2,*}$ and T. K. Jha $^{1}$} %MDPI: There are no explanations for dagger and ddagger. Please check and revise. Also, we removed Orcid because it was empty. Please confirm.
\AuthorNames{Prashant Thakur, Tuhin Malik and T. K. Jha}

\AuthorCitation{Thakur, P.; Malik, T.; Jha, T.K.}
\address{%
$^{1}$ \quad \textls[-20]{Department of Physics, {BITS-Pilani,} %MDPI: Please check if this is an institution. If so, please try to provide the complete name instead of the abbreviation and please make sure that the order is from subordinate to superior.
K. K. Birla Goa Campus, Goa 403726, India}\\
$^{2}$ \quad \textls[-20]{{CFisUC,} %MDPI: Please make sure that the order is from subordinate to superior. Also, if possible, pelase provide the complete name instead of the abbreviation. %TM: The complete name of the institute is Centro de Física da Universidade de Coimbra (CFisUC)
Department of Physics, University of Coimbra, {P-3004-516}~~Coimbra, Portugal %MDPI: Please check if Postal Code is correct. % TM: 3004-516 is correct. 
 }}

\corres{{Correspondence:} tuhin.malik@uc.pt } %MDPI: can you please provide us with the insittutional email?. Also, we revised the Corresponding author according to the ``*'' symbol. Please check and confirm or revise it. 

\abstract{Over the last few years, researchers have become increasingly interested in understanding how dark matter affects neutron stars, helping them to better understand complex astrophysical phenomena. In this paper, we delve deeper into this problem by using advanced machine learning techniques to find potential connections between dark matter and various neutron star characteristics. We employ Random Forest classifiers to analyze neutron star (NS) %MDPI: We removed green clolor in the paper. Please check and confirm. %TM: Confirmed 
properties and investigate whether these stars exhibit characteristics indicative of {dark matter} admixture. Our dataset includes 32,000 sequences of simulated NS properties, each described by mass, radius, and tidal deformability, {inferred} using recent observations and theoretical models. We explore a two-fluid model for the NS, incorporating separate equations of state for nucleonic and dark matter, with the latter considering a fermionic dark matter scenario. Our classifiers are trained and validated in a variety of feature sets, including the tidal deformability for various masses. The performance of these classifiers is rigorously assessed using confusion matrices, which reveal that NS with admixed dark matter can be identified with approximately 17\% {probability of misclassification as nuclear matter NS}. In particular, we find that additional tidal deformability data do not significantly improve the precision of our predictions. This article also delves into the potential of specific NS properties as indicators of the presence of dark matter. Radius measurements, especially at extreme mass values, emerge as particularly promising features. The insights gained from our study are pivotal for guiding future observational strategies and enhancing the detection capabilities of dark matter in NS. This study is the first to show that the radii of neutron stars at 1.4 and 2.07 solar masses, measured using NICER data from pulsars PSR J0030+0451 and PSR J0740+6620, strongly suggest that the presence of dark matter in a neutron star is more likely than only hadronic composition.}


\begin{document}

\section{Introduction}
{Neutron} 
stars (NS), because of their dense nature, are of great interest in the universe. Despite their importance in astrophysics, the insides of neutron stars remain a mystery. Neutron stars serve as natural laboratories for studying matter under extreme densities and pressures~\cite{1996cost.book.....G, book.Haensel2007, Rezzolla:2018jee}. Neutron star cores can reach densities approximately 4--5 times greater than normal nuclear saturation density, posing an intriguing astrophysical puzzle. The internal structure of neutron stars, particularly their cores, is one of the most mysterious topics in astrophysics. Despite advances in observational astronomy, current measurements of neutron star properties have not been able to definitively rule out the presence of exotic forms of matter, such as a deconfined quark phase or admixed dark matter. The extreme conditions within these stars make it possible for states of matter to exist that are not yet fully understood or observed under terrestrial conditions~\cite{Lattimer:2000nx}. Theoretical models suggest a range of scenarios for the core composition, from conventional nucleonic matter to more exotic forms, such as quark matter and strange matter, as discussed by Glendenning~\cite{1996cost.book.....G}. This uncertainty opens the door to considering the potential role of dark matter within the dense core of neutron stars~\cite{Kain:2021hpk,Das:2020ecp,Das:2021yny,Ruter:2023uzc}.

Dark matter, an invisible yet omnipresent part of the universe, is a major focus in astrophysics. It is estimated to make up 27\% of the mass--energy content of the universe, and its gravitational effects are seen in phenomena such as the rotational speeds of galaxies and the dynamics of galactic clusters~\cite{Rubin:1978kmz,Bauer:2017qwy,Salucci:2018hqu}. However, its mysterious nature and the difficulty in directly detecting it present a major challenge to our understanding of the cosmos. Neutron stars offer an exceptional and promising way to investigate the puzzle of dark matter. The intense gravitational fields of neutron stars make them a potential site for capturing dark matter particles~\cite{Bertone:2007ae,deLavallaz:2010wp,Guver:2012ba}, particularly Weakly Interacting Massive Particles (WIMPs), which are a leading candidate in {dark matter} models. The concept that dark matter could accumulate in the core of a neutron star and interact with its ordinary matter has generated considerable interest. Goldman and Nussinov (1989) first proposed this idea in their theoretical studies~\cite{Goldman:1989nd}, and Kouvaris further explored the implications of {dark matter} annihilation within neutron stars, suggesting potential observable signs~\cite{Kouvaris:2007ay,Kouvaris:2010vv}. The interaction between dark matter and neutron stars is not only about accumulation but also transformation. The dense environment within a neutron star could catalyze processes in which dark matter particles transition into different states~\cite{Bertone:2007ae,deLavallaz:2010wp,Guver:2012ba}. These transitions could affect the neutron star's temperature, it srotation, and even its life cycle. Moreover, the presence of dark matter within neutron stars could create innovative and exotic states of matter, which would challenge our current understanding of compact astrophysical objects. These theoretical possibilities have spurred a growing body of research aimed at using neutron stars as natural detectors for dark matter. Astrophysicists hope to detect signs indicative of {dark matter} interactions by analyzing the observational data of neutron stars, such as their mass--radius relationships, cooling rates, and magnetic-field structures. This effort not only increases our knowledge of neutron stars, but also potentially helps to solve the {dark matter} mystery that has long perplexed scientists~\cite{Raj:2017wrv,Goldman:1989nd,Gould:1989gw,Kouvaris:2007ay,Kouvaris:2010vv,deLavallaz:2010wp,Guver:2012ba,Ellis:2018bkr,Panotopoulos:2017idn,Das:2018frc,Das:2020vng,Tolos:2015qra,Shirke:2023ktu}.

The advanced LIGO and VIRGO collaborations have made a groundbreaking discovery of gravitational waves from the GW170817 event, which was a merger of two neutron stars. This, combined with multi-messenger observations of the binary, has opened up new possibilities for investigating the equation of state of dense matter and researching dark matter in compact objects~\cite{abbott2016observation,abbott2017gw170817}. In recent years, the scientific community has made considerable progress in understanding the interaction between dark matter (DM) and neutron stars (NS). This research has mainly focused on the effects of DM on the characteristics and detectability of neutron stars. For example, studies have examined the mass--radius relationships and the stability of DM-admixed neutron stars, with a comprehensive analysis of stability aspects~\cite{Leung:2011zz,Kain:2021hpk}. Additionally, refs.~\cite{Ellis:2018bkr,Das:2020ecp} have demonstrated that a DM core can significantly reduce the maximum mass of a neutron star. Ref.~\cite{Xiang:2013xwa} has extended these findings, discussing the formation of stable DM-admixed NSs and the potential variation in mass--radius relationships and the creation of a DM halo. This study has employed two equations of state for the neutron star's dense baryonic core, made up of piecewise generalized polytropes, and has investigated an asymmetric self-interacting fermionic dark matter component. The focus has been on different scenarios of admixed neutron stars under these specific conditions. Ref.~\cite{Emma:2022xjs} has taken this research further by simulating {dark matter} admixed neutron star binaries, showing that the presence of DM could shorten the merger remnant's lifespan and affect the brightness of electromagnetic signals. These collective studies have highlighted the potential impacts of DM in neutron stars, such as changes in maximum mass, tidal and surficial Love numbers, and mass--radius relationships. Research has also indicated that the inclusion of DM---bosonic or fermionic---might reduce the higher tidal deformability usually associated with a stiff nuclear equation of state (EOS). This phenomenon has been observed in both single-fluid and two-fluid approaches, as demonstrated in studies such as those by~\cite{Ivanytskyi:2019wxd, Rutherford:2022xeb,Karkevandi:2021ygv,Ellis:2018bkr}. Furthermore, to better understand the EOS of neutron stars, statistical Bayesian methods have been applied to analyze astrophysical observations, with nuclear matter parameters playing a key role. Although many studies have hypothesized the presence of DM in neutron stars, few have extensively explored the correlations between observable neutron star properties and dark matter parameters~\cite{Thakur:2023aqm}.

This research seeks to uncover the secrets of neutron stars and the impact of dark matter on their properties. Despite numerous studies, the universal correlations between neutron star properties in the presence and absence of dark matter remain unclear. To bridge this gap, our study explores these correlations, taking into account the uncertainties in nuclear EOS and dark matter EOS. We use machine learning classification algorithms that are trained on extensive datasets of neutron star properties, encompassing scenarios with and without dark matter, to investigate the feasibility of distinguishing neutron star characteristics based on the presence of dark matter. This machine learning endeavor has the potential to revolutionize our understanding of neutron stars. We also plan to determine the number of observations needed to conclusively classify the existence of dark matter within neutron stars. This could provide a new way to view and understand these enigmatic celestial bodies. Our research combines traditional astrophysical methods with cutting-edge machine learning techniques to venture into uncharted territories of neutron star and dark matter research.

The paper has the following organization. In Section~\ref{sec2}, we introduce the basic formalism of the equation of state for nuclear matter and dark matter. In Section~\ref{results}, we present and discuss the results of the current study. Finally, in Section~\ref{conclusions}, we provide concluding remarks.

\section{The Equation of State (EOS)}\label{sec2}
{To calculate the neutron star properties with admixed dark matter configurations, we use a two-fluid approach~\cite{Das:2020ecp,Kain:2021hpk,Rutherford:2022xeb,Karkevandi:2021ygv}. In this approach, separate EOSs are required for two~distinct fluids that interact only gravitationally, as described by two-fluid equations. For the nuclear matter component, the EOS is based on only nucleonic degrees of freedom, considering beta-equilibrated cold dense matter. For the dark matter component, we consider a fermionic particle with vector interactions only. Both EOSs are treated within the relativistic mean field (RMF) framework. It should be noted that our limited understanding of dark matter and its minimal constraints allow the validity of various types of modeling, including those involving bosonic particles~\cite{Rutherford:2022xeb,Giangrandi:2022wht,Karkevandi:2021ygv}. Despite this diversity, recent observational constraints do not definitively rule out any particular model.  In the following section, we will explain the methodology used to construct both EOS.}

\subsection{Nuclear Matter EOS}\label{SubsecB1}
We use the nuclear EOS dataset created in ref.~\cite{Malik:2023mnx}, using the RMF approach, {which only includes nucleonic degrees of freedom}. {{Specifically}, {\it ``Set 0''} %MDPI: Please check if italic is necessary. %TM: Yes, it is necessary. 
in the above-referenced article denotes the full dataset, which covers the entire range of prior.} This approach is a mean-field theory that incorporates non-linear meson terms, both self-interactions and mixed terms. Here, a Bayesian inference method is employed to fine-tune the model parameters, guided by constraints such as nuclear saturation properties, the requirement that neutron stars exceed a maximum mass of 2 M$_\odot$, and the low-density pure neutron matter EOS as determined by N3LO calculations in chiral effective field theory. This collection of 16,000~equations of state (EOS) is included in the dataset. {The Lagrangian density that accounts for the baryonic components is expressed as follows:}

\vspace{-6pt}\begin{equation}
  \mathcal{L}=   \mathcal{L}_{\rm N}+ \mathcal{L}_{\rm M}+ \mathcal{L}_{\rm NL}
\end{equation} 
with
\begin{equation}
\begin{aligned}
\mathcal{L}_{\rm N}=& \bar{\Psi}\Big[\gamma^{\mu}\left(i \partial_{\mu}-g_{\omega} \omega_{\mu}-%\frac{1}{2}
g_{\varrho} {\boldsymbol{t}} \cdot \boldsymbol{\varrho}_{\mu}\right) \\
&-\left(m-g_{\sigma} \sigma\right)\Big] \Psi \\
\mathcal{L}_{\rm M}=& \frac{1}{2}\left[\partial_{\mu} \sigma \partial^{\mu} \sigma-m_{\sigma}^{2} \sigma^{2} \right] \\
&-\frac{1}{4} F_{\mu \nu}^{(\omega)} F^{(\omega) \mu \nu} 
+\frac{1}{2}m_{\omega}^{2} \omega_{\mu} \omega^{\mu} \nonumber\\
&-\frac{1}{4} \boldsymbol{F}_{\mu \nu}^{(\varrho)} \cdot \boldsymbol{F}^{(\varrho) \mu \nu} 
+ \frac{1}{2} m_{\varrho}^{2} \boldsymbol{\varrho}_{\mu} \cdot \boldsymbol{\varrho}^{\mu}.\\
    			\mathcal{L}_{\rm NL}=&-\frac{1}{3} b ~m ~g_\sigma^3 \sigma^{3}-\frac{1}{4} c g_\sigma^4 \sigma^{4}+\frac{\xi}{4!} g_{\omega}^4 (\omega_{\mu}\omega^{\mu})^{2} \nonumber\\&+\Lambda_{\omega}g_{\varrho}^{2}\boldsymbol{\varrho}_{\mu} \cdot \boldsymbol{\varrho}^{\mu} g_{\omega}^{2}\omega_{\mu}\omega^{\mu}.
\end{aligned}
\label{lagrangian}
\end{equation}

\textls[-20]{The Dirac spinor $\Psi$ stands for the nucleon doublet (consisting of neutron and proton) with a bare mass $m$. The couplings of the nucleons to the meson fields $\sigma$, $\omega$, and $\varrho$ are denoted by \linebreak  $g_{\sigma}$, $g_{\omega}$, and $g_{\varrho}$ respectively, with their respective masses being $m_\sigma$, $m_\omega$, and $m_\varrho$. The parameters $b$, $c$, $\xi$, and $\Lambda_{\omega}$, which control the strength of the non-linear terms, are determined in conjunction with the couplings $g_i$ (where $i=\sigma, \omega, \varrho$) by imposing a set of conditions.}

\subsection{Dark Matter EOS}\label{SubsecB2}
We consider a simple model for the dark matter equation of state that is similar to the Lagrangian of the nuclear model, as employed in ref.~\cite{Thakur:2023aqm}. This model has a single fermionic component ($\chi_D$) and a dark vector meson $V_D^{\mu}$ that couples to the conserved DM current through $g_{v d}\bar{\chi}_D\gamma_{\mu}\chi_DV^{\mu}_D$. The Lagrangian and the corresponding EOS in the mean-field approximation are expressed as follows:
\begin{eqnarray}
   {\cal{L}}_{\chi} &=&  \bar{\chi}_D \left[\gamma_{\mu}(i\partial^{\mu} - g_{v d}V^{\mu}_D) - m_{\chi}\right]\chi_D \nonumber\\ 
               &-& \frac{1}{4}V_{\mu\nu,D}V_D^{\mu\nu} + \frac{1}{2}m_{v d}^2V_{\mu,D}V_D^{\mu}
\end{eqnarray}
\vspace{-12pt}
\begin{eqnarray}
\varepsilon_{\chi} = \frac{1}{\pi^2}\int_{0}^{k_D} dk~k^2\sqrt{k^2 + m_\chi^2} + \frac{1}{2} c_{\omega}^2\rho^2_D 
\end{eqnarray}
\vspace{-12pt}
\begin{eqnarray}
P_{\chi} = \frac{1}{3\pi^2}\int_{0}^{k_D} dk \frac{k^4}{\sqrt{k^2 + m_{\chi}^2}} +\frac{1}{2} c_{\omega}^2\rho^2_D   
\end{eqnarray}

The ratio of $g_{vd }$ to $m_{vd }$ is denoted as $c_{\omega }$, and the bare mass of the fermionic dark matter is represented by $m_{\chi}$. These two factors, in addition to the dark matter Fermi momenta, are what determine the dark matter equation of state. 
{The density of the fermionic dark matter is represented by $\rho_D$, which is linked to the mean-field value of the 'dark vector' meson.} The Fermi momenta of the dark matter are what determine the amount of dark matter density/mass that is accumulated inside neutron stars. The characteristics of neutron stars that have been mixed with dark matter depend on the dark matter equation of state and the mass fraction of the dark matter. 

\textls[-20]{Our approach to analyzing the structure of NSs with an admixed dark matter configuration with a two-fluid approach involved using four different random nucleonic EOS with varying stiffness to account for uncertainties in the nuclear EOS domain. We also employed a range of {dark matter} EOS developed using the relativistic mean-field (RMF) approach. \textls[-20]{For each nuclear matter scenario, we examined a comprehensive set of 4000~unique combinations of dark matter parameters that gives NS maximum mass above {1.9~$M_\odot$, which is a value within $3\sigma$~of the PSR J0348+0432 mass, 2.01$~\pm~0.04\, M_\odot$.} These parameters ($c_{\omega}$, $m_{\chi}$, and $F_{\chi}$)} were selected from uniform distributions within the specified ranges described in Table~\ref{tab:dmt}.
Notably, the $F_{\chi}$ is define a dark matter mass fraction
$
F_{\chi} = \frac{M_{\chi}(R_\chi)}{M{(R)}}
$.
Here, $M_{\chi}(R_{\chi})=4\pi\int_0^{R_\chi} r^2\varepsilon_{\chi} (r) dr$ represents the total accumulated dark matter gravitational mass within $R_{\chi}$, where the dark matter pressure reaches zero. In general, we consider 16,000 admixed configurations of the sequence of NS properties, such as mass, radius, and tidal deformability. This sequence represents the entire curve between mass, radius, and tidal deformability.}

\begin{table}[H]
\centering
\caption{{The varied} range of dark matter model parameters.}\label{tab:dmt} %MDPI: Figures and Table were moved after their first citations. Please check and confirm. %TM: OK
\setlength{\tabcolsep}{11.0pt}
\newcolumntype{C}{>{\centering\arraybackslash}X}
\begin{tabularx}{\textwidth}{CCCCCC}
\toprule
\multicolumn{2}{c}{\boldmath{$m_{\chi}$}} & \multicolumn{2}{c}{\boldmath{$c_{\omega}$}} & \multicolumn{2}{c}{\boldmath{$F_{\chi}$}} \\
\multicolumn{2}{c}{\textbf{GeV}}        & \multicolumn{2}{c}{\textbf{fm}} %TM: fm will be in small 
& \multicolumn{2}{c}{\textbf{\%}}        \\ \midrule
\textbf{Min}           & \textbf{Max}            & \textbf{Min}             & \textbf{Max}            & \textbf{Min}           & \textbf{Max}          \\ \midrule
0.5           & 4.5           & 0.1               & 5             & 0             & 25         \\ \bottomrule
\end{tabularx}
\end{table}

\textls[-20]{We investigate dark matter with a mass range of 0.5--4.5 GeV~\cite{Calmet:2020pub}, a self-interaction measure between 0.1 and 5 fm~\cite{Xiang:2013xwa}, and a mass fraction of up to 25\%~\cite{Ciancarella:2020msu}. 
{For the nucleon--dark matter interaction within neutron stars, estimates suggest a dark matter capture rate around $10^{25}$ GeV/s for dark matter mass near 1 GeV~\cite{Bell:2020jou}. Over the neutron star's lifetime, this implies a modest accumulation of dark matter, suggesting that alternative mechanisms might be necessary. For instance, dark matter production during supernovae events~\cite{DeRocco:2019jti} or the conversion of neutrons into dark matter particles could contribute significantly to dark matter content in neutron stars, as explored in the context of dark matter admixed neutron stars~\cite{Ellis:2018bkr}.} }

\subsection{The Datasets}
\label{dataset}
Our study encompasses a comprehensive collection of 32,000 (32~K) neutron star (NS) properties, including mass--radius curves and mass--tidal deformability curves. Half of these (16,000, or 16~K) are derived from models with dark matter, and the remaining 16~K are from models without dark matter. To facilitate the machine learning classification task, we structured our datasets as follows.

\textls[-20]{We constructed a feature vector \(X \) comprising five randomly chosen masses, uniformly distributed between 1 solar mass and the maximum mass specified by each curve.~Similarly~to these five masses, we included five radii and five tidal deformabilities in the feature vector.~Thus, each {vector} \(X \) consisted of 15 elements in total: 5 masses, 5 radii, and 5 tidal~deformabilities.} %MDPI: Values are not consistent in the paper. Please check and revise. E.g., X is normal here and in italic before. Same for other values too (e.g., y, M0, etc.) and same for the rest of the paper. %TM: changed 

For the target vector \(Y \), we assigned a value of zero to the curves that represented nuclear-matter configurations only and a value of one to those that included {dark matter} configurations. Consequently, our target vector \(Y \) was a binary indicator, with zero signifying the absence of dark matter and one indicating its presence in the model. This binary classification system effectively distinguished between neutron star models with and without {dark matter} influence.

\section{Results} \label{results}
In this section, we aim to examine the observational properties of neutron stars. Our analysis focuses on results obtained using a varied collection of nuclear matter EOS without dark matter, as well as a range of dark matter EOS. In the two fluid approach, four~different nuclear EOS are considered, each with a different stiffness. 

\begin{figure} [H]
 \includegraphics[width=0.98\linewidth]{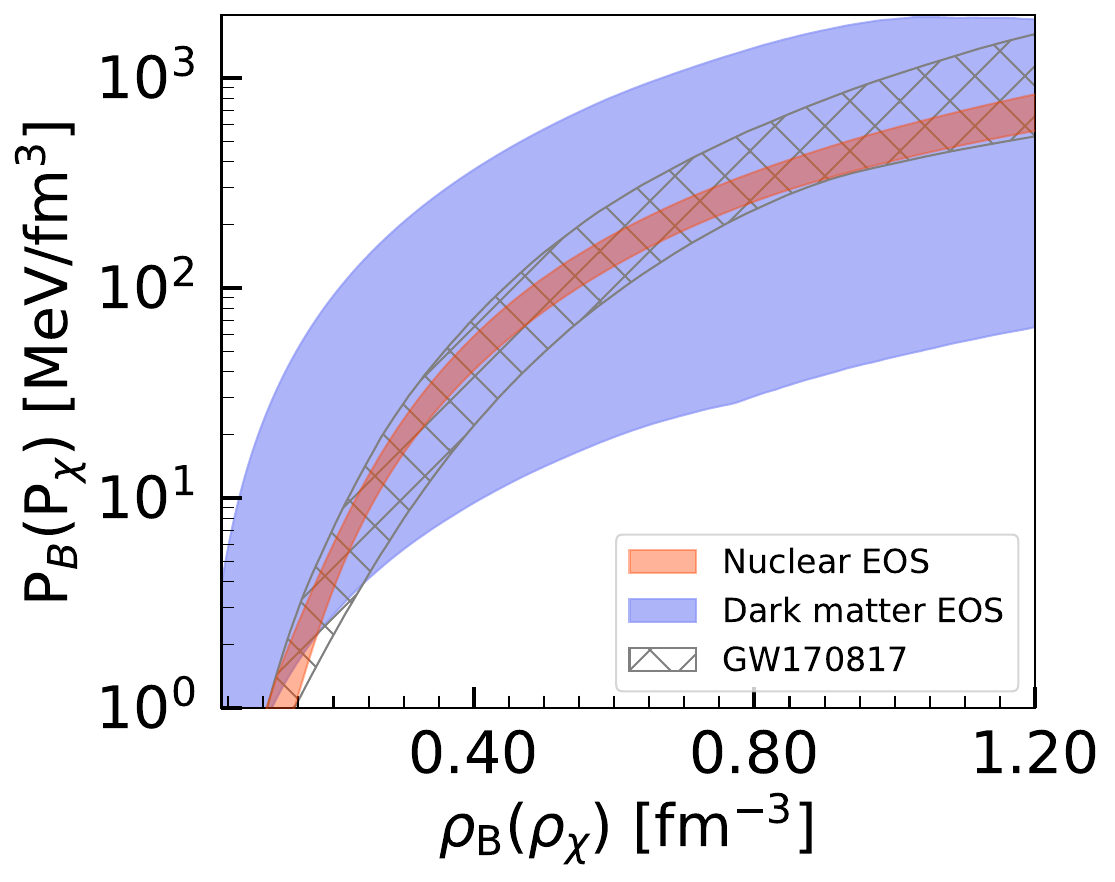}
 \caption{{The} 90\% confidence interval of the dark matter (nuclear matter) equation of state is represented by the shaded blue (light orange) area. This is the pressure $P_{\chi}$ ($P_B$) as a function of density $\rho_{\chi}$ ($\rho_B$). The hatched grey band is the prediction from the GW170817 event.} %MDPI: Please make sure that values are consistent with the main text. %TM: Yes, it is consistent. 
 \label{fig:eos_dark_matter}
 \end{figure}

In Figure~\ref{fig:eos_dark_matter}, we show the equation of state for both nuclear matter (highlighted in orange) and dark matter (highlighted in blue). The nuclear matter equation of state was obtained with minimal constraints on nuclear saturation properties, neutron star masses above two solar masses, and low density constraints from chiEFT theory~\cite{Malik:2023mnx}. The dark matter equation of state was derived by varying the model parameters within the range given in Table~\ref{tab:dmt}. This range is wider than that for nuclear matter, as there are fewer constraints on dark matter equations of state. Additionally, the graph displays the domain of equations of state derived from GW170817, represented by the cross-hatched overlay. It should be noted that this domain is only applicable to nuclear matter equations of state, as it was derived using a single-fluid approach.

\begin{figure}[H]
     %\centering
     \includegraphics[width=0.98\linewidth]{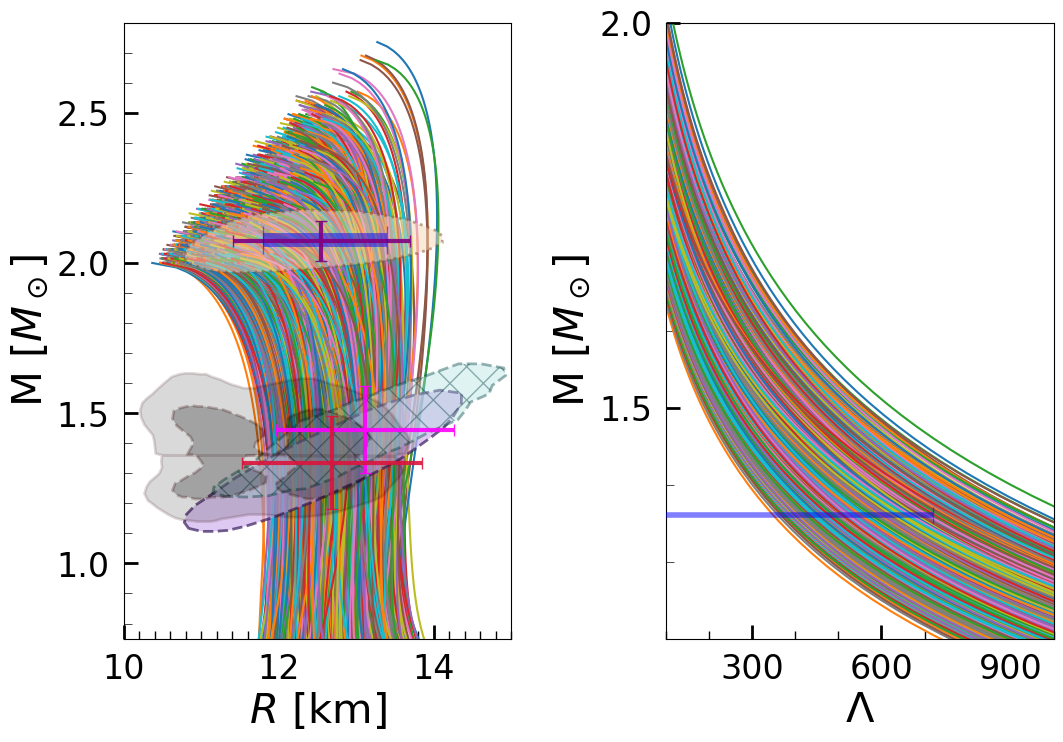}
     \caption{{The mass}--radius and mass--tidal deformability curves for the 16,000 nuclear matter equation of state, i.e., the properties of neutron stars without dark matter, are shown, respectively. The gray zones in the left panel indicate the 90\% (solid) and 50\% (dashed) confidence intervals for the binary components of the GW170817 event~\cite{LIGOScientific:2018hze}. The cyan and yellow zones in the same panel represent the $1\sigma$ (68\%) credible area of the 2-D posterior distribution in the mass--radius domain from millisecond pulsar PSR J0030+0451~\cite{Riley:2019yda,Miller:2019cac}, while the violet zone is from PSR J0740+6620~\cite{Riley:2021pdl,Miller:2021qha}, both derived from NICER X-ray data. The horizontal (radius) and vertical (mass) error bars reflect the $1\sigma$ credible interval derived from the same NICER data's 1-D marginalized posterior distribution. The blue bars depict the radius of PSR J0740+6620 at 2.08~$M_\odot$ (left panel) and its tidal deformability at 1.36 $M_\odot$ (right panel)~\cite{LIGOScientific:2018cki}.} %MDPI: Also, please make sure that all colors are explained.
     %MDPI: We recommend using (a) and (b) notation for subfigures instead of left and right. Please chekc and revise if necessary. % TM: Confirmed
     \label{fig:mrl_nm}
 \end{figure}

Figure~\ref{fig:mrl_nm} presents the relationship between mass, radius, and tidal deformability for neutron stars, based on a set of 16,000 equations of state that describe nuclear matter without the influence of dark matter. In the mass--radius plot on the left, the gray areas represent the confidence intervals (CI) for the binary components observed in the GW170817 event, with solid lines indicating a 90\% CI and dashed lines indicating a 50\% CI. Additionally, we have incorporated data from NICER's X-ray observations of the pulsars PSR J0030+0451 and PSR~J0740+6620. For PSR~J0030+0451, the cyan and yellow shaded regions correspond to the $1\sigma$ (68\%) confidence zone from its 2-D posterior distribution in the mass--radius space. The violet shaded region pertains to PSR J0740+6620. The error bars displayed—horizontal for radius and vertical for mass---represent the credible intervals $1\sigma$ from the marginalized posterior distributions of 1D derived from the same NICER data.
In the right panel, which plots the mass against the tidal deformability, the blue bars mark the specific measurements for PSR J0740+6620: its radius at a mass of 2.08~$M_\odot$ and its tidal deformability at 1.36 $M_\odot$. From the figure, it is evident that the nuclear matter EOS set we have used encompasses a wide array of neutron star properties. The maximum mass of these neutron stars ranges between 2 and 2.75 solar masses ($M_\odot$), indicating a wide spread in both the radius and tidal deformability values across the set, demonstrating the versatility of the EOS employed. Crucially, our results show good alignment with the various astrophysical constraints derived from recent observations.

\begin{figure}[H]
     %\centering
     \includegraphics[width=0.98\linewidth]{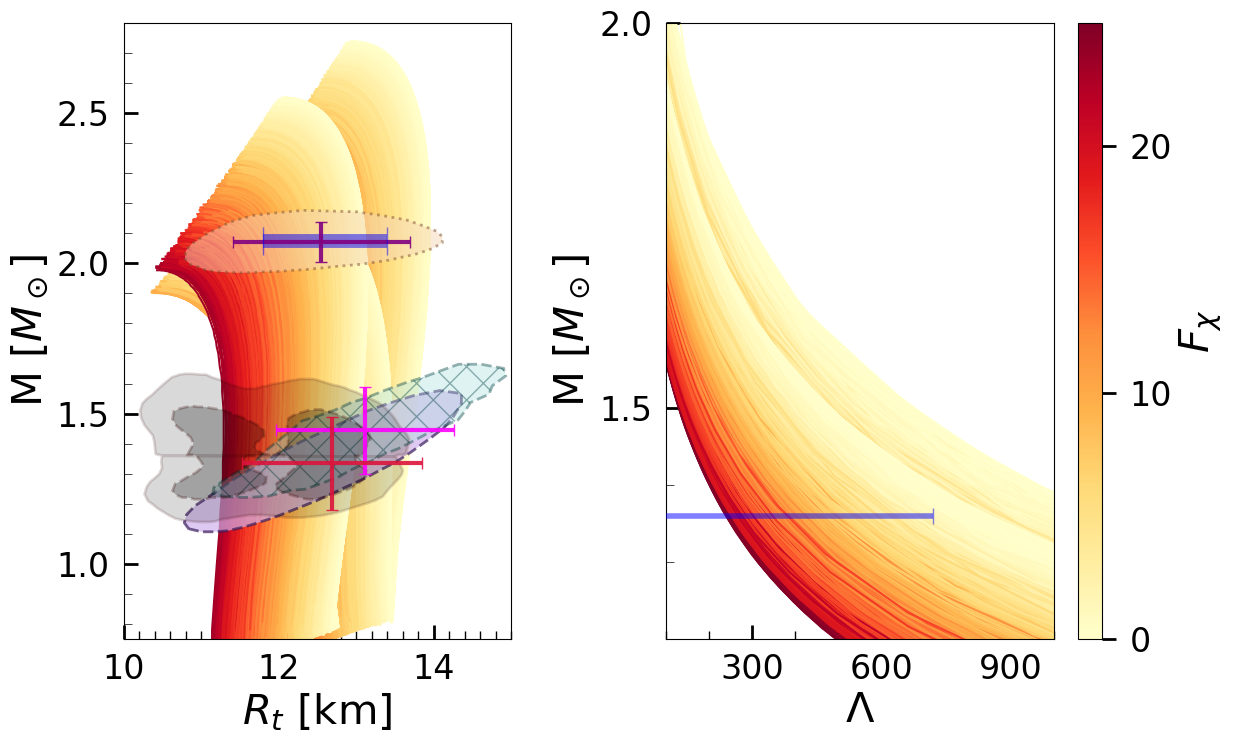}
     \caption{{Same} as Figure~\ref{fig:mrl_nm}, but the properties of neutron stars with a dark matter configuration of 16,000 EOS are depicted. The fraction of admixed dark matter over the total mass of the neutron star, $F_\chi$, is also shown. {see text for details}. The range of colors, starting from yellow and progressing to dark red, indicates the varying mass fraction of dark matter, ranging from zero to the maximum percentage shown.} %MDPI: Please explain all lines and colors. %TM: Added a line. Now it is OK. 
     \label{fig:mrl_dm}
 \end{figure}

\textls[-20]{The figure referenced as Figure~\ref{fig:mrl_dm} illustrates the mass--radius (left panel) and mass--tidal deformability (right panel) relationships for neutron stars, based on 16,000 equations of state  that include dark matter. These properties have been computed using a two-fluid} approach, as previously described. The nuclear matter component in the two-fluid models is represented by four randomly selected EOS from the aforementioned set of 16,000 nuclear matter EOS. These four were specifically chosen for their varying stiffness, effectively spanning the current range of uncertainty within the nuclear EOS. It is important to note that the left panel of the figure presents the variation of the neutron star mass in relation to the total radius ($R_t$). This total radius ($R_t$) is defined as the point at which the pressure of both fluids, nuclear matter and dark matter, disappears. In this context, \(F_\chi \) represents the fraction of the total neutron star mass that is constituted by dark matter. An increase in \(F_\chi \) results in a corresponding decrease in both the mass and the radius of the neutron star, as shown in the left panel. Similarly, the right panel shows that dimensionless tidal deformability also decreases with an increase in \(F_\chi \). The study cited as~\cite{Thakur:2023aqm} concluded that \(F_\chi \) is strongly correlated with neutron star properties at any given mass, provided the nuclear matter EOS remains constant. However, when variations in nuclear matter EOS are introduced, effectively incorporating the uncertainty in the nuclear EOS, the correlations with \(F_\chi \)~dissipate. This finding underscores the need to disentangle the effects of uncertainties stemming from both nuclear EOS and dark matter on the properties of neutron stars.

\begin{figure}[H]
    \includegraphics[width=0.98\linewidth]{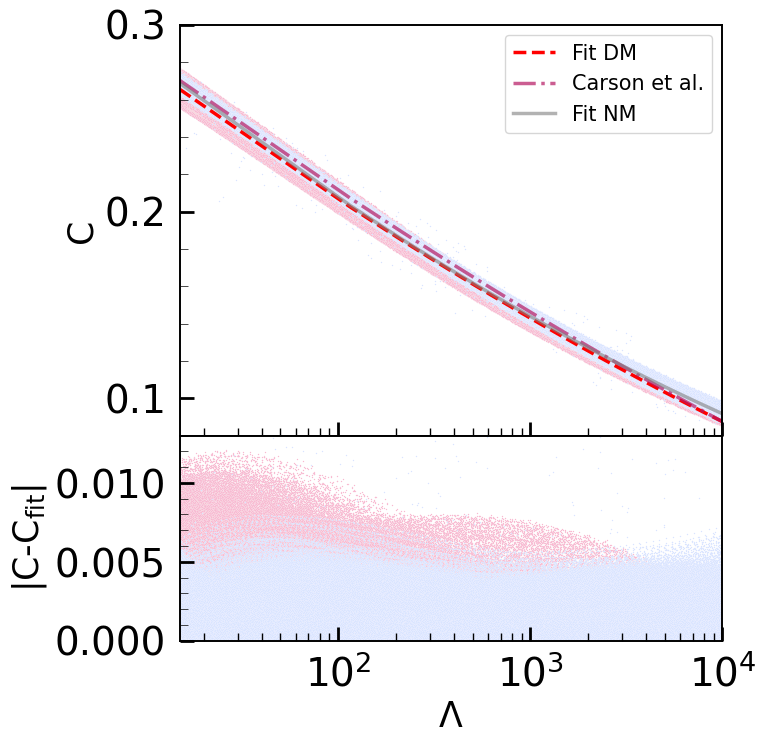}
    \caption{{The figure} %MDPI: Please change the hyphen (-) into minus sign ($-$, "U+2212"). e.g., "-1" should be "$-$1". Also, please make sure that values are consistent with the main text.
 illustrates the C--Love universal relation for EOS involving only nuclear matter (depicted in light blue) and those that include dark matter (shown in light pink). The red line indicates the fit for the dark matter EOS, while the gray line shows the fit for the nuclear matter EOS, both fitted using Equation~\eqref{unieq}. The lower panel of the figure displays the residuals from these fittings. Additionally, the figure includes a comparison with the results from ref.~\cite{Carson:2019rjx}, specifically focusing on the single-fluid Tolman--Oppenheimer--Volkoff (TOV) scenario without dark matter, which is highlighted in a pink-red color.}
    \label{fig:lovec}
\end{figure}

{Figure}~\ref{fig:lovec} examines the robustness of certain universal relations in the context of neutron star properties, particularly when dark matter is present. Previous research has suggested that some of these relations may not be valid when quark matter is taken into account. Therefore, our investigation focuses on the \(C-\Lambda \) relation, which has been found to be unaffected by the equation of state when only nucleonic degrees of freedom are considered, to determine the impact of admixed dark matter. The C--Love relationship is depicted in {Figure}~\ref{fig:lovec} with light pink scatter plots for nuclear EOS sets without dark matter and light blue scatter plots for those with admixed fermionic dark matter. We used a simple polynomial curve, as described by the equation below, to fit the EOS data:
\vspace{-6pt}\begin{equation}
    \label{unieq}
    C = \sum^2_{k=0} a_k (\ln{\Lambda})^k.
\end{equation}

{The fitting process for the nuclear EOS data yielded coefficients \(a_0=0.36935858 \pm 0.00001225\), \(a_1=-0.04011723 \pm 0.00000397\), and \(a_2=0.0010823 \pm 0.0000003\). In comparison, the admixed dark matter EOS set resulted in \(a_0 = 0.36054566 \pm 0.00002011\), \(a_1 = -0.0375908 \pm 0.00000931\), and \(a_2 = 0.00086283 \pm 0.00000093\).} The lower panel in Figure~\ref{fig:lovec} presents the absolute differences from these fits, revealing a deviation of approximately 1\% that diminishes to 0.5\% for higher values of \(\Lambda \), with the nuclear matter EOS case consistently close to 0.5\%. Our findings were juxtaposed with the results from Figure 4 in ref.~\cite{Carson:2019rjx}, which reported on a constrained EOS with coefficients \(a_0 = 0.3617 \), \(a_1 = -0.03548 \), and \(a_2 = 0.0006194 \). The constrained EOS exhibited an absolute difference of less than 1\%, whereas the unconstrained EOS showed around a 1\% deviation. From this comparative analysis, we infer that dark matter does not significantly disrupt the universal \(C-\Lambda \) relation, suggesting that the relationship between compactness \(C \) and tidal deformability \(\Lambda \) remains stable even when dark matter is considered within the NS structure.

\begin{figure}[H]
    \includegraphics[width=0.98\linewidth]{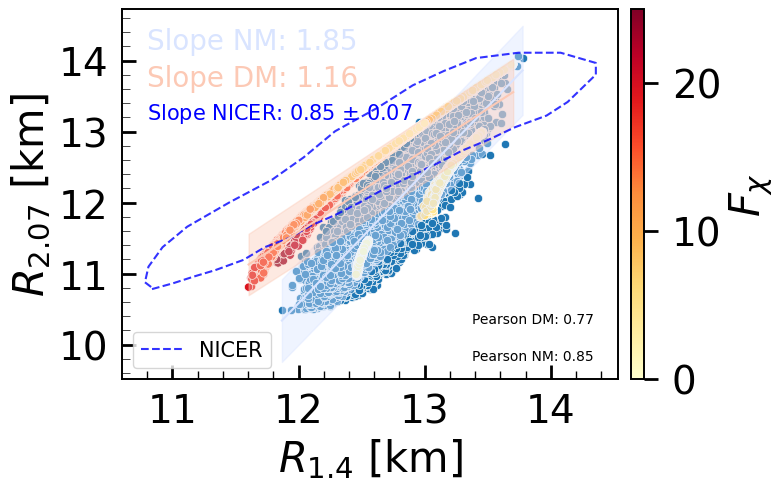}
    \caption{The figure displays the relationship between the radii of neutron stars with masses of 1.4~and 2.07~solar masses (M$_\odot$). The comparison is made between two different scenarios: one with dark matter and one without. The radius in the scenario that includes dark matter is the total radius of the neutron star, including the dark matter component.}
    \label{fig:r_cor}
\end{figure}

In our analysis, depicted in Figure~\ref{fig:r_cor}, we explore the correlation between the radii of neutron stars with masses of 1.4 and 2.07 solar masses (M$_\odot$), comparing scenarios with and without dark matter. Inspired by a recent study~\cite{Lin:2023cbo}, which presents a method to detect phase transitions in neutron stars by examining the radius correlations, including NICER's observations of PSR J0740+6620 and PSR 0030+0451, we extend this approach to investigate the presence of dark matter. We plotted the relationship between the radii of neutron stars with masses of 1.4 and 2.07 solar masses (M$_\odot$), comparing scenarios with and without dark matter. In the {dark matter} scenario, the total radius of the neutron star includes the dark matter component. We found that the Pearson correlation coefficient is 0.77 for the dark matter set and 0.85 for the nuclear matter set, indicating a slight weakening of the correlation with dark matter. Furthermore, we compared the constraints derived from NICER in blue, which were obtained by marginalizing the 1$\sigma$ posteriors of the NICER measurements of these two pulsars over the NS mass. Then, the radius data of these two~observations yielded this blue domain. This analysis considered only the data analyzed by the Riley et al. group. Significantly, we observed that the slope of the relationship between $R_{2.07}$ and $R_{1.4}$ undergoes a considerable change in the presence of dark matter. The slope is 1.85 in the absence of dark matter and 1.16 with admixed dark matter. The line fitted to each scenario is plotted along with the uncertainty band of $1 \sigma$. The presence of dark matter appears to reduce this slope. In particular, the slope for the NICER constraints is $0.85 \pm 0.07$, showing substantial overlap with the dark matter admixed set. This overlap suggests that the NICER measurements of the radii of PSR~J0740+6620 and PSR 0030+0451 might indicate a higher likelihood of the presence of dark matter compared to solely hadronic matter.

\textls[-20]{The research conducted thus far has strongly encouraged us to explore the potential for dark matter to be revealed in the observational properties of neutron stars, such as mass, radius, and dimensionless tidal deformability, through the use of machine learning. The machine learning tools are becoming increasingly important in neutron star physics~\cite{Ferreira:2021pni, Carvalho:2023ele, Vidana:2022prf, Thete:2023aej, Soma:2023rmq, Anil:2020lch}. We use the {robust} dataset of neutron star properties of both nuclear matter and admixed dark matter configurations that were previously discussed. Our goal is to create a classification system that can distinguish between the presence and absence of dark matter in neutron~stars.}

We create two features, denoted as $X1$ and $X2$. Feature $X1$ consists of mass and radius values, while $X2$ includes an additional element---tidal deformability. These values are generated through random uniform sampling, with the range for mass and radius extending from 1 to their respective maximum values. The construction of $X1$~involves a total of 10~elements, each comprising 5~values of mass and radius. On the other hand, $X2$ has an added dimension of 5 tidal deformability measurements, bringing its total length to 15 elements. Section~\ref{dataset} discusses structuring the dataset for machine learning training. It should be noted that the target vector $Y$ has only one element, which is set to zero for $X$ with only nuclear matter and 1 for admixed dark matter. 

In our study, we employ the Random Forest classifier~\cite{breiman2001random}, recognized as one of the most effective classification algorithms, to analyze our data. To begin, we combine two datasets, each containing 16,000 entries, ensuring a thorough shuffle to guarantee uniformity in the data distribution. Then, this merged dataset is divided into three distinct parts: 60\% for training, 20\% for validation, and the remaining 20\% reserved for testing.

The training set, which constitutes 60\% of the data, is used to teach the Random Forest model the patterns and relationships that are inherent to the data. The validation set, comprising~20\%~of the data, plays a critical role in fine-tuning the model. During this phase, we adjust the hyperparameters of the Random Forest classifier. Hyperparameters are the configuration settings used to structure the learning process and can significantly impact the performance of the model. By tweaking these parameters while observing the model's performance on the validation set, we aim to find the optimal configuration that yields the best results. This process of hyperparameter tuning is essential to prevent issues like overfitting, where the model performs well on the training data but poorly on unseen data. The final 20\% of the data, which are discarded as a test dataset, are used to evaluate the performance of the fully trained model. This dataset is crucial, as it represents new unseen data for the model, providing a realistic assessment of how the model will perform in real-world scenarios. It is important to note that these test data are not involved in the training or validation process. By keeping these data separate, we ensure that our evaluation of the model's effectiveness is unbiased and indicative of its true predictive capabilities.

\begin{figure}[H]
    \includegraphics[width=0.45\linewidth]{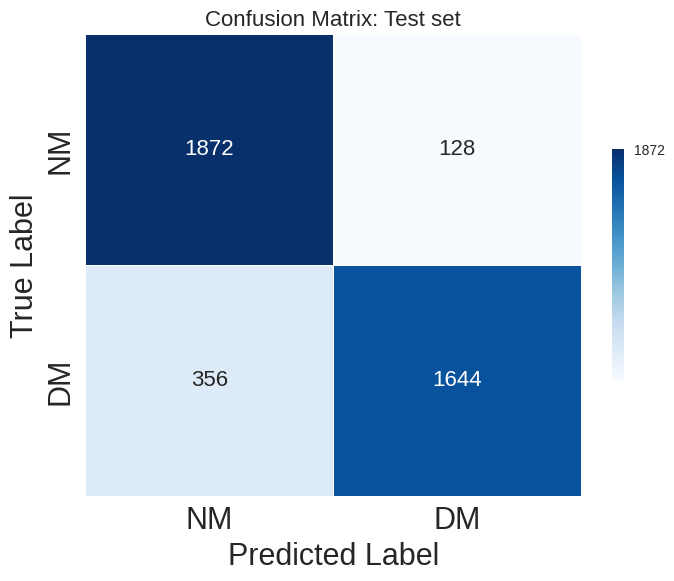}
    \includegraphics[width=0.45\linewidth]{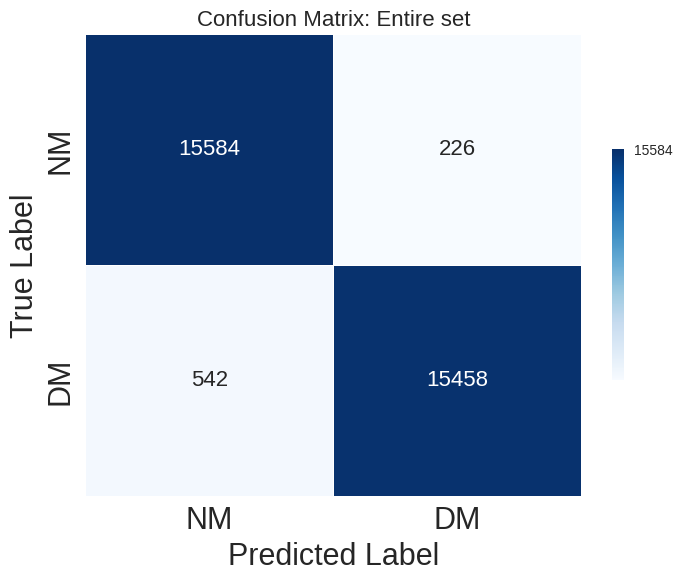}
    \caption{{The confusion} %MDPI: Please use commas to separate thousands for numbers with five or more digits (not for four digits) in the picture. e.g., "10000" should be "10,000" . #TM: The number format indicated in the picture is fine. Please go ahead with this. 
    matrices displayed are for the test set and the entire set, respectively (model trained with $X1$ feature, i.e., NS mass and radius).} %MDPI: Please try to use (a) and (b) notations for subfigures instead of left and right. %TM: Changed 
    \label{fig:confusionx1}
\end{figure}

\begin{figure}[H]
    \includegraphics[width=0.45\linewidth]{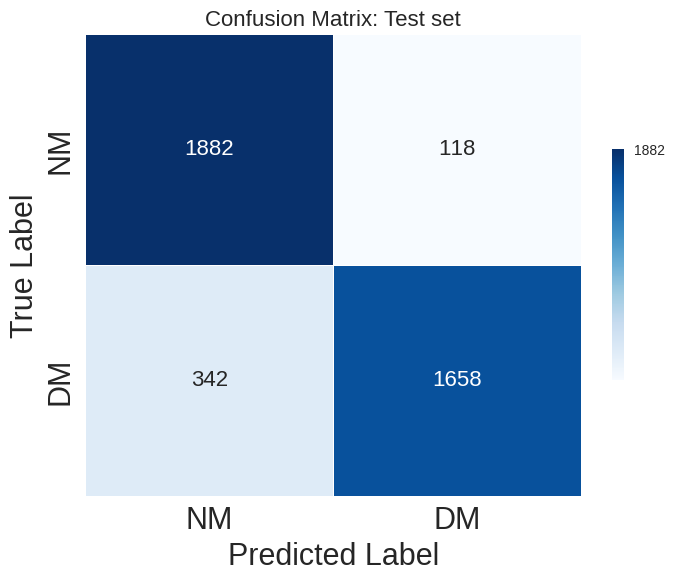}
    \includegraphics[width=0.45\linewidth]{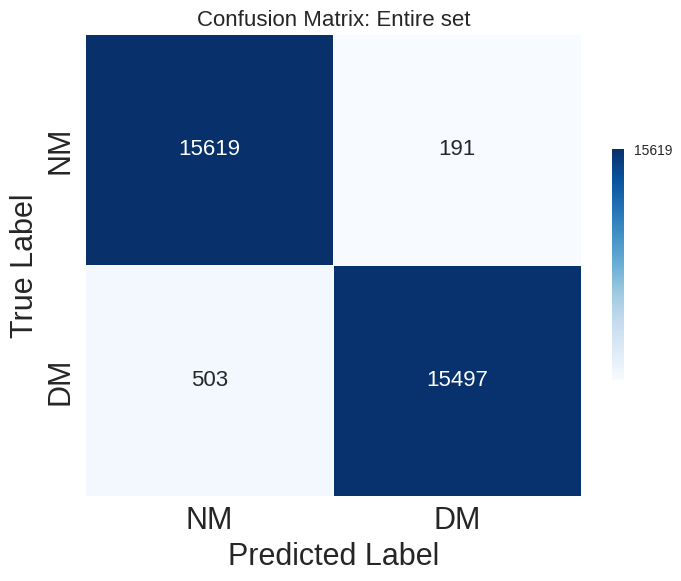}
    \caption{\textls[-20]{{Same as} shown in Figure~\ref{fig:confusionx1}, but with the trained model on the $X2$ feature; it has five additional features compared to the previous, which are the tidal deformability for five different~masses}.} %MDPI: Please use commas to separate thousands for numbers with five or more digits (not for four digits) in the picture. e.g., "10000" should be "10,000" #TM: The number format indicated in the picture is fine. Please go ahead with this. 
    \label{fig:confusionx2}
\end{figure}

In Figure~\ref{fig:confusionx1}, two panels represent confusion matrices for a Random Forest classifier applied to train model with only the $X1$ feature. The left panel shows the prediction of the trained model on the test set and the right panel shows the entire dataset. A confusion matrix is a table that is often used to describe the performance of a classification model on a set of data for which the true values are known. It cross-tabulates the actual class labels with the predicted class labels, providing insight into the accuracy and types of errors made by the classifier.
In the left panel, the confusion matrix for the test set reveals that the classifier accurately predicted NS with only nuclear matter (NM) 1872 times and with admixed dark matter (DM) 1644 times. However, there were instances of misclassification, indicated by off-diagonal numbers: 128 instances of NM were incorrectly classified as DM and 356~instances of DM were incorrectly classified as NM. These errors highlight the instances where the classifier was challenged to distinguish between the two classes. Moving to the right panel, the confusion matrix for the entire dataset shows that the classifier correctly identified NM 15,584 times and DM 15,458 times. Misclassifications are also present here, with NM being mistaken for DM 226 times and vice versa 542 times. The percentage of false positive for NS with only nuclear matter (NM) is 6.4\% (1.98\%) and NS with admixed dark matter (DM) is 17.8\% (4.98\%) for the set of tests (entire). It should be noted that the test set contains data separated from those not involved in training, but the entire set contains data involved in training. Figure~\ref{fig:confusionx2} shows a confusion matrix for the model trained in the $X2$ feature set, which includes five additional features: tidal deformability for five different mass categories. The purpose of introducing these additional features was to determine whether the inclusion of tidal deformability constraints could improve the accuracy of the predictions. \textls[-20]{The results show that the rate of false positives, incorrectly predicting NM as {dark matter} DM, is 5.9\% for the test set and 1.7\% for the entire dataset. On the other hand, the rate of false positives for neutron stars with admixed dark matter is 17.1\% for the test set and 4.69\% for the entire set. These figures suggest that the inclusion of tidal deformability data does not significantly enhance the predictive power of the model.}

\unskip

\begin{figure}[H]
    \includegraphics[width=0.8\linewidth]{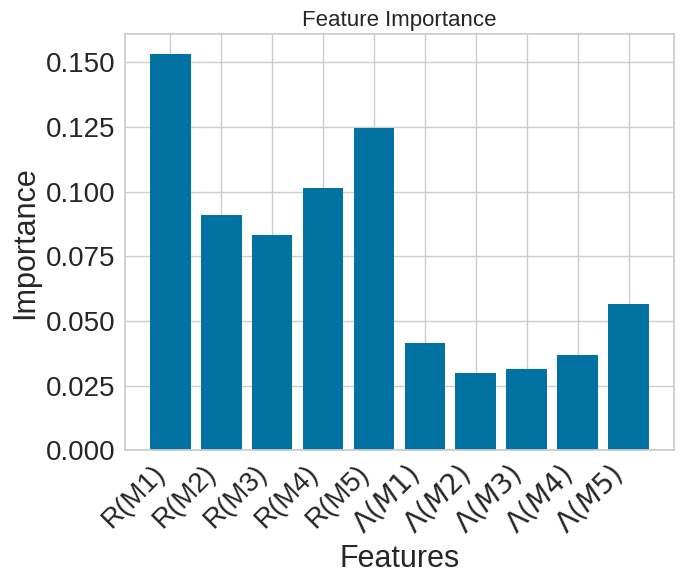}
    \caption{Feature importance for Random Forest classification from $X2$.}
    \label{fig:enter-label}
\end{figure}

As Random Forest is a tree-based model, one can also extract the importance of different features (or the mean decrease in impurity) within a trained model {in predicting the target variable.} {This is accomplished by monitoring the reduction in model precision when each feature is taken out of the model. To put it more simply, it shows the impact of each feature on the model's decision-making process, with higher numbers indicating a greater effect on the classification result.} In Figure~\ref{fig:enter-label}, we plot the importance of the feature that delineates the varying significance of the features employed by the Random Forest model in the context of classifying the existence of dark matter within neutron stars (NS). Radius measurements at various mass points (`$R(M1)$' through `$R(M5)$') dominate the feature importance, with the plot revealing that `$R(M1)$', the radius at a lower mass point, holds the greatest predictive power, followed by `$R(M4)$', a higher mass point. This pattern suggests that radius measurements, especially at the extremes of the mass range considered, are critical for the model to discern the presence of dark matter in the NS. Tidal deformability features (`$\Lambda(M1)$' through `$\Lambda(M5)$') are also included, although they exhibit less influence on the model's decisions. This insight underscores the necessity of precise radius measurements over a range of masses as a more determinant factor in revealing or classifying dark matter within NS. The comparative lower importance of the tidal deformability suggests that, while useful, they do not contribute as significantly to the model's classification ability as the radius measurements.

\section{Conclusions} \label{conclusions}
In this article, we explored the combination of dark matter and neutron stars through a two-fluid model, concentrating solely on a fermionic dark matter equation of state. To replicate this situation, we carefully employed EOS for nucleonic matter and dark matter within a relativistic mean-field (RMF) approach, which included a comprehensive collection of 16,000 different EOS. The nuclear EOS was the result of Bayesian inference with minimal constraints on nuclear saturation properties, an NS maximum mass of more than 2 solar masses, and a low-density pure neutron matter constraint. For the dark matter component, a straightforward model was adopted, featuring a singular fermionic entity interacting with a dark vector meson. The resulting {dark matter} EOS was contingent on the interaction strength with the dark vector meson and the intrinsic mass of the fermionic dark matter. The interplay between the {dark matter} EOS and the proportion of dark matter present critically determined the resultant properties of the neutron stars.

A comprehensive study was carried out on the combination of neutron stars and dark matter, using four different nucleonic equations of state and a range of dark matter equations of state. This research looked at 32,000 neutron stars properties curve, examining their mass, radius, and tidal deformability. The study also delved into the integrity of universal relations in the presence of dark matter, affirming the stability of the compactness--tidal deformability (C{--}$\Lambda$) relation. %MDPI: Please check if this should be en dash or minus sign. %ENGLISH: en dash is correct %TM: dash is perfect. It is now fine. 
{Furthermore, we investigate the correlation between the radii $R_{1.4}$ and $R_{2.07}$ for neutron stars with masses 1.4 and 2.07 $M_\odot$, respectively.} The two radii are closely related, but the inclusion of dark matter in neutron stars can alter the gradient. For the first time, our analysis has taken advantage of the most recent observational data from NICER, which strongly suggests that dark matter must be included in neutron star models. As our knowledge advances and our observational powers increase, the intricate part that dark matter plays in the universe will become more distinct, giving us new perspectives on the essential character of our cosmos.

\textls[-15]{We employed machine learning techniques, particularly Random Forest classifiers, to classify neutron stars (NS) in the presence of dark matter, based on their inherent properties. We evaluated the predictive accuracy of these classifiers, trained on various feature sets, using confusion matrices. Our findings suggest that the properties of NS with a mixed dark matter configuration can be identified with approximately 17\% {probability of misclassification as nuclear matter NS}. Adding constraints related to tidal deformability does not significantly reduce this uncertainty. This figure reflects both the inherent uncertainty of the classifier model and the variability in NS properties. Future research should include a more focused analysis to distinguish between these sources of uncertainty, as well as a broader examination using a variety of machine learning algorithms. Additionally, our analysis sought to identify critical features that would be useful for future observations to detect dark matter within NS. The results suggest that measurements of radius at both lower and higher masses appear to be very promising indicators for the presence of dark~matter.}

\vspace{6pt} 
%%%%%%%%%%%%%%%%%%%%%%%%%%%%%%%%%%%%%%%%%%
\authorcontributions{P.T. was responsible for the numerical simulation in this work. T.M. and T.K.J. contributed to the formulation of the problem and analysis of the results. All the authors collaborated in the writing and editing of the manuscript. {All authors} %MDPI: Newly added information. Please check and confirm.
have read and agreed to the published version of the manuscript.
}

\funding{{This work} was partially funded by FCT (Fundação para a Ciência e a Tecnologia, I.P, Portugal) through Projects No. UIDP/\-04564/\-2020, No. UIDB/\-04564/\-2020, and 2022.06460.PTDC. T.M. would like to thank FCT---Fundação para a Ciência e a Tecnologia for the support provided through Project No. EXPL/FIS-AST/0735/2021. %MDPI: ``This research received no external funding'' or ``This research was funded by NAME OF FUNDER grant number XXX.'' and  and ``The APC was funded by XXX''. Check carefully that the details given are accurate and use the standard spelling of funding agency names at \url{https://search.crossref.org/funding}, any errors may affect your future funding.
}

 \dataavailability{This article did not generate any new data and instead utilized a dataset that was previously published elsewhere. Nevertheless, if the author requests it, we are able to supply our structured dataset for their research.
 %MDPI: We encourage all authors of articles published in MDPI journals to share their research data. In this section, please provide details regarding where data supporting reported results can be found, including links to publicly archived datasets analyzed or generated during the study. Where no new data were created, or where data is unavailable due to privacy or ethical restrictions, a statement is still required. Suggested Data Availability Statements are available in section ``MDPI Research Data Policies'' at \url{https://www.mdpi.com/ethics}.
 } 

% Only for journal Nursing Reports
%\publicinvolvement{Please describe how the public (patients, consumers, carers) were involved in the research. Consider reporting against the GRIPP2 (Guidance for Reporting Involvement of Patients and the Public) checklist. If the public were not involved in any aspect of the research add: ``No public involvement in any aspect of this research''.}

% Only for journal Nursing Reports
%\guidelinesstandards{Please add a statement indicating which reporting guideline was used when drafting the report. For example, ``This manuscript was drafted against the XXX (the full name of reporting guidelines and citation) for XXX (type of research) research''. A complete list of reporting guidelines can be accessed via the equator network: \url{https://www.equator-network.org/}.}

\acknowledgments{ The author, P.T, would like to acknowledge CFisUC, University of Coimbra, for their hospitality and local support during his visit in May - June 2023 for the purpose of conducting part of this research. The authors are grateful to the Laboratory for Advanced Computing at the University of Coimbra for providing {HPC} resources that have contributed to the research results reported in this paper, URL: \hyperlink{https://www.uc.pt/lca}{https://www.uc.pt/lca} ({accessed on 1 Jan 2024} %MDPI: Please add the access date (Format: Date Month Year). e.g., (accessed on 1 January 2020).
).}

\conflictsofinterest{The authors declare no conflicts of interest. %MDPI: Declare conflicts of interest or state ``The authors declare no conflicts of interest.'' Authors must identify and declare any personal circumstances or interest that may be perceived as inappropriately influencing the representation or interpretation of reported research results. Any role of the funders in the design of the study; in the collection, analyses or interpretation of data; in the writing of the manuscript; or in the decision to publish the results must be declared in this section. If there is no role, please state ``The funders had no role in the design of the study; in the collection, analyses, or interpretation of data; in the writing of the manuscript; or in the decision to publish the results''.
} 

 \begin{adjustwidth}{-\extralength}{0cm}
% %\printendnotes[custom] % Un-comment to print a list of endnotes

\reftitle{References}

\PublishersNote{}
\end{adjustwidth}
\end{document}